# Terahertz metallic photonic crystals integrated with dielectric waveguides

**Dejun Liu[1, *], Lin Chen[2], Feng Liu[1, *]**

[1]*Department of Physics, Shanghai Normal University, Shanghai 200234, China*
[2]*Shanghai Key Lab of Modern Optical System, University of Shanghai for Science and Technology, Shanghai 200093, China*
*\*dejunliu1990@gamil.com; fliu@shnu.edu.cn*

**Abstract:** Compact and low-loss photonic crystal waveguides are critical in integrated terahertz (THz) applications. Compared with pure metal or dielectric photonic crystal waveguides, hybrid (metal/dielectric) integrated waveguides provide a simple way to further improve the field confinement and the propagation loss. In this work, we investigate a hybrid waveguide consisting of metallic photonic crystals and dielectric films in 0.1-1.0 THz. Photonic crystal waveguides based on metal pillar arrays (MPAs) support two resonance modes including the fundamental and high-order transverse magnetic (TM) modes and then form one apparent bandgap in 0.45-0.55 THz. The high-order TM-mode shows higher confinement than the fundamental mode and are thus sensitive to the dielectric film on the MPAs. The propagation loss and field confinement can be optimized by changing the dielectric film's thickness and refractive index. The investigation shows that the lowest loss is less than 0.1 cm$^{-1}$ at 0.68 THz because the high-order TM-mode THz waves are tightly confined inside the hybrid waveguide. This work proves that such hybrid waveguides based on metallic photonic crystals are promising to develop as a compact integrated terahertz device.

**Keywords:** Terahertz integrated waveguides, metallic photonic crystals, hybrid waveguides, low-loss, high field confinement.

## 1. Introduction

Terahertz (0.1 THz to 10 THz) applications such as communication, imaging, and sensing arouse increasing attention in recent years [1-3]. However, bulky optical THz systems need precise alignment and servicing. Furthermore, guided THz wave propagation is also one of the greatest challenges because of the strong absorption from water [4]. A compact waveguide that extended the THz pulse propagation distance is therefore highly desired in the integrated THz systems. Pure dielectric or metal photonic crystal waveguides are beneficial towards this end, as they have been used to achieve low propagation loss [5-10]. On the basis of solid dielectric waveguides, dielectric photonic crystal waveguides with a number of air holes have been proposed, which can suppress the THz field in the subwavelength scale [5]. The previous result shows that the propagation loss is proportional to $\lambda^2/a^3$, where a is the core radius [11]. The propagation loss can be effectively reduced with a few orders of magnitude by increasing the core diameter. A porous polyethylene (PE) waveguide shows lower modal losses (~0.05 cm$^{-1}$) than a solid rod fiber with the same diameter (~0.15 cm$^{-1}$) [12]. Such waveguides cannot neglect the material absorption from the dielectric. Metallic waveguides are considered as a great candidate for waves guiding because the metal shows low material absorption in THz region [13-19]. A simple bare metallic wire was reported to transport THz pulses, which realizes a loss of 0.03 cm$^{-1}$ around 0.25 THz [14]. In general, there is a trade-off between mode confinement and propagation loss. Loss is inversely proportional to the mode confinement [17-18]. For instance, metallic parallel plate waveguide (PPWG) offers high field confinement but suffers from giant power loss [17]. An efficient approach to further reduce the THz propagation loss and improve mode confinement is to combine a metallic waveguide with a dielectric waveguide. The formed novel waveguide calls hybrid waveguide, which scheme thereby breaks through the diffraction limitation on deep-subwavelength confinement, but much extends propagation lengths [20-25]. A hybrid metal wire–dielectric terahertz waveguide supporting tightly confined, air-bound modes at both high and low frequencies has been proposed [25]. This metal wire-dielectric waveguide well addresses the issue of pure metal wire waveguide that suffers from a loosely confined mode. Metallic photonic crystal-based waveguides are promising to achieve narrow field confinement and low loss due to the supporting surface plasmon polaritons (SPPs) [26-29]. A photonic crystal waveguide based on a metal rod array has been demonstrated, which achieves a loss lower than 0.1 cm$^{-1}$ [27-29]. However, the field confinement and propagation loss of hybrid waveguides based on metallic photonic crystals have not yet been explored.

In this work, a hybrid waveguide is investigated in numerical at 0.1-1 THz, which consists of metal pillar arrays (MPAs) and a dielectric waveguide. Two resonance modes including the fundamental and high-order transverse magnetic (TM) modes are found in the transmission spectrum of MPAs, which is separated by one apparent bandgap in 0.41–0.55 THz. We firstly investigate the propagation loss of MPAs by changing the pillar length. Results demonstrated that longer pillar is beneficial to reduce the propagation loss. To further reduce THz waves propagation loss, a low-loss dielectric film is attached to the MPAs. By changing the dielectric film's thickness and refractive index, not only the propagation loss but also the field confinement can be optimized. The investigation shows the lowest loss is less than 0.1 cm$^{-1}$ at 0.68 THz due to the high-order TM-mode THz waves are strictly confined inside the hybrid waveguide. The investigation shows this hybrid waveguide is suitable to develop as terahertz integrated components because of the controllable field confinement and propagation loss.

## 2. Metal pillar array (MPA)-based hybrid waveguides

Figure 1 presents the configuration of the proposed metallic pillar arrays (MPAs) with dielectric waveguides in which the MPAs are considered as metallic photonic crystals [29]. Geometrical parameters are denoted in Fig. 1, where the metal pillars with 0.12 mm diameters (D) are periodically arranged on the substrate along the X- and Y-axes with period Λ=0.30 mm. The dimensions of the pillar diameter and length are supposedly uniform to interact with THz waves. The dielectric film works as a THz ribbon dielectric waveguide, fully covering the MPA with various thicknesses and refractive indices. In the finite-difference-time-domain (FDTD) calculation, the material of pillars is considered as the perfect electronic conductor (PEC). The input wavefronts belong to the Gaussian beams and the electric field of the input transverse magnetic (TM) waves is polarized perpendicular to the metal pillars.

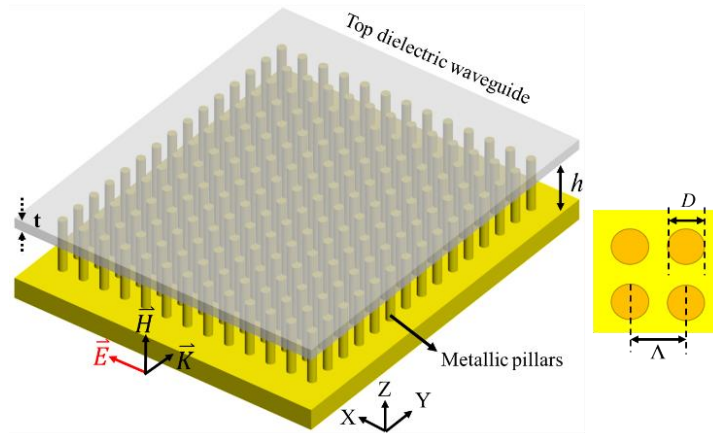

Fig.1. (a) Configuration of the hybrid waveguide based on metal pillar arrays and dielectric waveguides.

## 3. Results and discussion

### 3.1 Waveguide modes in MPA waveguides

Figures 2 (a) shows the schematic of one-line MPA (30 layers), which is initially presented to reveal the propagations of TM waveguide modes. The period ($\Lambda$), diameter (D), and the height (*h*) of metal pillars are 0.30, 0.12, and 1 mm, respectively. The transmission spectrum of TM modes for the one-line MPA shows in Fig. 2 (b), where insets are electric field distributions of 0.44 THz and 0.68 THz at the output of MPAs. The features of TM wave transmission for one-line MPA is the low-frequency pass spectrum, where the spectral peak is at 0.44 THz. As the frequency increases, the transmittance of the high-frequency band increases. The transmittance difference between the low and high-frequency bands originates from the existence of the waveguide mode. The MPA shows high confinement at the frequency of 0.44 THz, where the fields can be accumulated on the metal rod surface. However, the field of 0.68 THz is weaker, which is only located at the pillar tip. Figures 2 (c) and (d) show the electric field distributions at the top of MPAs in the X-Y plane. The electric field distribution of 0.44 THz is different from that of 0.68 THz. The waveguide mode of 0.44 THz propagates along the metal pillars, which can be regarded as confined modes. However, the electric field of 0.2 THz is obviously decayed by the MPA layer, which is radiated from the MPA with the increase of layers.

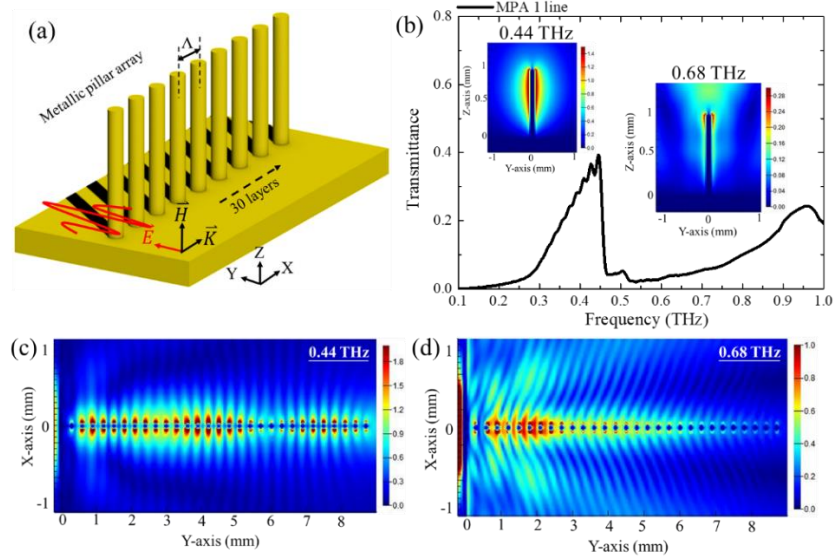

Fig. 2 (a) Schematic of the one-line MPA. (b) The transmittance spectrum of one-line MPA, Inset is the electric field distribution of 0.44 and 0.68 THz in the Z-Y plane. (c) and (d) are the electric field distribution of 0.44 THz and 0.68 THz in the X-Y plane, respectively.

We further investigated the incident TM waves to interact with the other multiple lines MPAs. Here, the additional 14 lines MPA periodically integrated with the X-axial interspace of 0.3 mm-period. Figure 3 (a) shows the transmittance spectra of MPAs with different pillar lengths. For the 1.0-mm height MPA (the cyan line), an obviously complete bandgap from 0.45 to 0.55 THz between the first and second transmission bands can be observed in the spectrum of TM modes. The bandgap is contributed by the additional 14 lines MPAs. The center frequency of bandgap is 0.50 THz, which is satisfied with the principle of Bragg frequency calculated from $f_B=C/2n\Lambda$, where *n* is the effective refractive index (*n=1* here) and $\Lambda$ is the period of pillar arrays [30]. Thus, waves with frequencies lying within this band, also called photonic bandgap, behave in a manner akin to total internal reflection. Furthermore, the high-frequency band is enhanced by the multi-line MPAs. The results also indicate that TM waves above 0.90 THz are not guided through the MPAs because of the high waveguide resistance [29]. The length of pillar arrays has a giant effect on the transmittance of TM modes. For the waveguide modes is lower than 0.45 THz, the spectral transmittance clearly increases with metal pillar length. In other words, the waveguide mode confinement is improved by the higher pillar length of MPAs. For instance, in the low-frequency band, the 0.6 mm-MPA has a transmittance of 0.08 at 0.4 THz, but the 1.0 mm-MPA has a transmittance of 0.52 at 0.35 THz. Similarly, the transmittance of the high-frequency band is also improved by the increment of pillar length.

To further understand the waveguide mode properties of MPAs, we calculated the dispersion relation of 2D MPAs in the first Brillouin zone, results are shown in Fig. 3 (b). In the calculation, we only consider two modes for MPAs because the transmittance of TM modes is extremely low when the frequency larger than 0.9 THz. For the fundamental TM mode (mode 1) below the light line, the propagation constant grows monotonically, indicating that the MPA behaves as a conventional waveguide. In the frequency range of 0.42-0.55 THz, corresponding to the photonic bandgap, THz waves cannot pass through the MPA and Bragg reflection occurs. The high-order mode is mode 2 locating in the frequency range of 0.55-0.90 THz, corresponding to the high-frequency band. Based on the previous studies, above the Bragg bandgap, the Bloch–Floquet mode is a leaky mode and the optical waves are partially radiated out of the waveguide [30]. Insets in Fig. 3 (b) are the field distribution of TM eigenmodes in the X-Y plane, corresponding to fundamental and high-order modes. The THz field of the fundamental mode is confined at the metal pillar surface. Inversely, the high-order mode field is distributed around the metal pillar. These results mean that high-order modes show lower confinement than that of fundamental modes. However, in the MPA case in vertical directions, the high-order mode above the Bragg bandgap is primarily confined in MPA waveguides, which have been proved by experimental and simulation results [28-29].

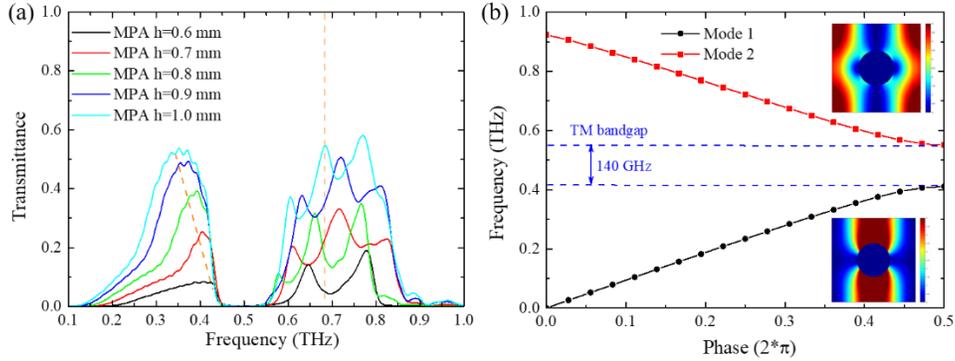

Fig. 3 (a) Transmission spectra of TM modes for 13 lines MPA with different pillar lengths.
(b) Schematic dispersion diagram in the first Brillouin zone of 2D MPAs for TM modes

We consider 20 MPA layers (30/50 layers=0.6 mm) as one waveguide interval and analyze the waveguide propagation loss of MPAs with different pillar lengths. The waveguide propagation losses are obtained from the equation *[ln(Tr.$_1$/Tr.$_2$)]/[2(L$_2$−L$_1$)]*, where $L_{1,\,2}$ indicates two waveguide lengths for different numbers of MPA layer and $Tr._{1,\,2}$ indicates the related transmittance at different numbers of the MPA layer [30]. Figure 4 (a) shows that the waveguide propagation loss of MPAs obviously varies with the changes of pillar length because of the variation of mode confinement. In the low-frequency band, the waveguide loss of h=0.6 mm-MPA at 0.35 THz is less than 0.1 cm$^{-1}$. The increased pillar length reduces the propagation loss of MPAs when the frequency is lower than 0.3 THz. For the high-frequency band, the propagation loss is efficiently reduced by the longer pillar MPAs. For example, at 0.68 THz, the 0.6 mm- and 1.0 mm-MPA have a loss of 2 cm$^{-1}$ and 0.2 cm$^{-1}$, respectively. Figure 4 (b) shows the electric field distribution of 0.68 THz at the output of MPAs. Pillar lengths such as

0.6, 0.7, 0.9, 1.0 mm are selected as examples to investigate, where the color bar is normalized to the same value of 1.0. As shown, the field of 0.68 THz is confined at the center of MPA and the field is weak when the pillar length is 0.6 mm. With the increasing of pillar length, the field in MPAs is significantly enhanced. It means that longer pillars are efficiently trapped the THz waves in MPAs and achieves lower propagation loss. The propagation loss of MPA-based waveguides can also be further reduced by optimizing the air-gap size between metal pillars, which has been studied in our previous works [29].

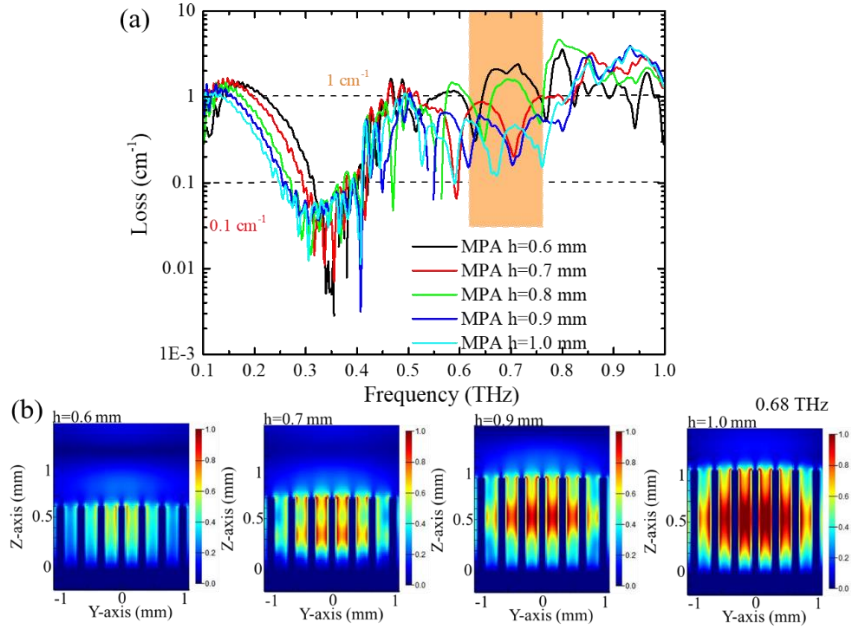

Fig. 4 (a) The propagation loss of MPAs with different pillar length. (b) The electric field distribution of 0.6, 0.7, 0.9, and 1.0 mm-MPA at 0.68 THz.

### *3.2 Waveguide modes in MPA-based hybrid waveguides*

### *3.2.1 Propagation loss dependent on the dielectric film thickness*

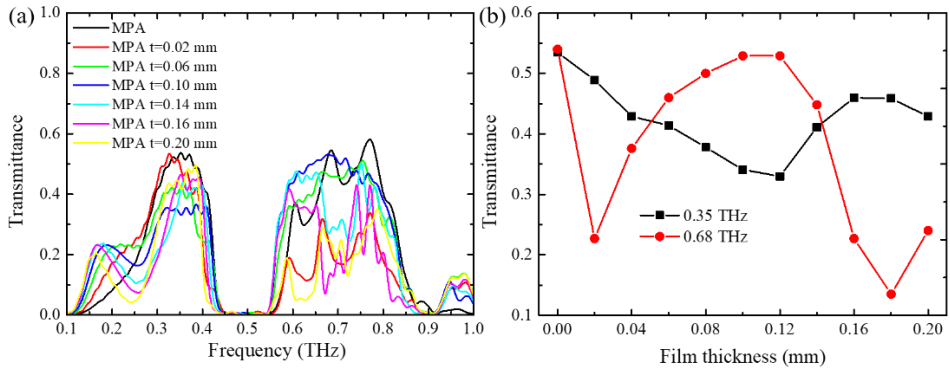

Fig. 5 (a) The calculated transmittance spectra of MPAs integrated with different thickness dielectric films. (b) The transmittance of 0.35 THz and 0.68 THz with the changes in film thickness.

Here, we investigate the transmission properties of MPA-based hybrid waveguides. Figure 5 (a) illustrates the transmittance spectra of MPA-based hybrid waveguides for various dielectric film thicknesses. The thin film in FDTD is assumed as a dielectric film with a refractive index of 1.6 (loss=0.01 cm$^{-1}$). Apparently, such a dielectric film integration does not giant change the MPA spectral feature of the photonic bandgap. In addition, the spectral peak in the low- and high-frequency bands show no obvious red-shift for the variation of film thickness. However, the transmittance values at the low- and high-frequency bands have been modified, which is compared with that of the blank MPA. The variation of the spectral transmittance due to the thickness variation of the thin-film integration can be used for THz sensing. We summarize the transmittance of 0.35 and 0.68 THz versus the thickness of dielectric films. As shown in Fig. 5 (b), the transmittance of 0.35 THz decreases as the film thickness increases, which is inversely proportional to the increment of film thicknesses. When the film thickness larger than 0.12 mm, the related transmittance increases. This result indicates that the fundamental mode is indeed sensitive to the film thickness changes. At 0.68 THz, the transmittance increases and then decreases with the thickness of the dielectric film. For example, the attached 0.02 mm-film efficiently reduces the transmittance from 0.54 to 0.21. As the film thickness increases from 0.02 mm to 0.12 mm, the transmittance is up to 0.53. The different response between low- and high-frequency band peak is highly correlated to the Z-axial field distribution. As discussed in previous works [29], the mode field at the low-frequency band is almost confined at the top of MPAs, in contrast to the field at high-frequency band inside the MPA. The obvious changes in spectral transmittance are mainly caused by the large interaction volume between the THz waves and film.

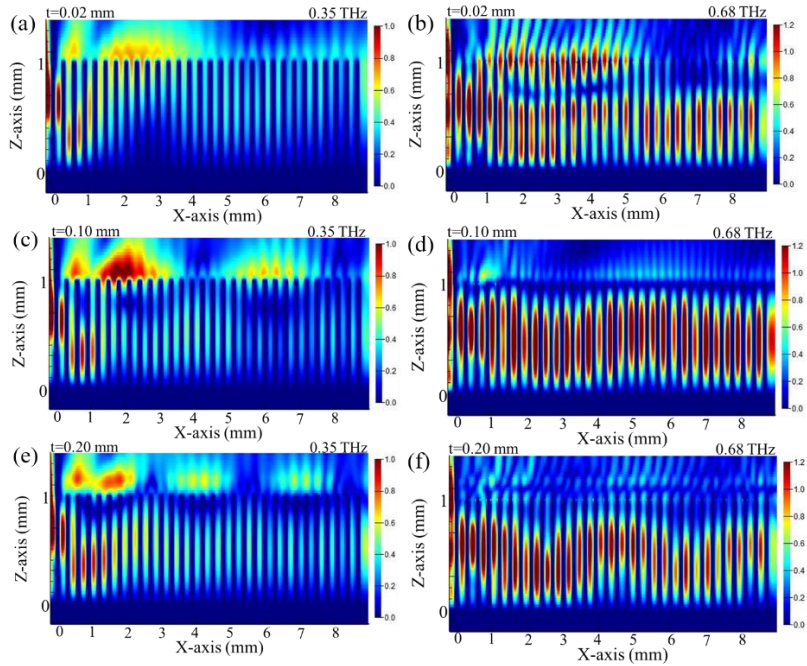

Fig. 6 Electric field distribution at 0.35 THz and 0.68 THz for different film thickness, such as 0.02, 0.10, and 0.20 mm.

To further understand the principle of THz waves propagation in MPA-based hybrid waveguides, we simulated the electric field distribution at 0.35 THz and 0.68 THz of 0.02, 0.10, and 0.20 mm-thick film-based waveguides. In Fig. 6, the maximum magnitude of the electric field is normalized to the same value. The fundamental and high-order modes exhibit different modal field distribution, such as surface and confined modes, respectively [29]. As given in Fig. 6 (a), the electric field distribution in the Z-X plane, the 0.35 THz waves is outside the hybrid waveguide for attaching one 0.02 mm film. Compared with the mode at 0.35 THz, the high-order mode at 0.68 THz exhibits different field distributions. As shown in Fig. 6 (b), the THz wave has been divided into two parts. One-part THz wave is coupled into the 0.02 mm-thick film and then propagating along the film waveguide. After propagated for a certain length, this wave becomes evanescent field and finally leaked into the air space. Another part of the THz wave is trapped inside the hybrid waveguides. Thus, the 0.02 mm thick film results in a loosely confined mode and then causes low transmittance at 0.68 THz. For the 0.10 mm-thick film, 0.35 THz field is modified by the thicker dielectric waveguide. In Fig. 6 (c), the 0.10 mm-film can guide a part of THz fields. For the cross-field distribution of the Z-X plane at 0.68 THz in Fig. 6 (d), compared with that of the blank MPA, the integrated dielectric film raises lateral confinement to involve more fraction of field inside the MPA. The 0.20 mm-film improves the confinement of 0.35 THz in contrast to 0.10 mm-film. However, at 0.68 THz, the MPA integrated with a 0.20 mm-film shows low transmittance. The thick film on the top of MPAs results in a strong reflection in the MPA; thus, the optical path length has been extended (Fig. 6 (f)).

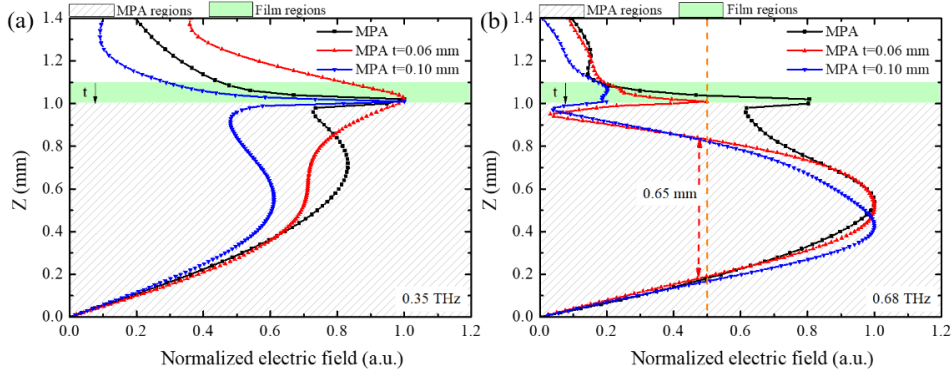

Fig. 7 Electric field distributions along the Z-axis at the output of hybrid waveguide at 0.35 THz (a) and 0.68 THz (b).

Now, we look at the modal field confinement when MPAs integrated with a dielectric film waveguide. Figure 7 shows the results of the modal field distribution at 0.35 THz and 0.68 THz. The maximum field confined at the interface between MPAs and air space when the frequency is 0.35 THz, which is shown in Fig. 7 (a) (black symbol line). The strongest field of 0.35 THz waves becomes more confined to the MPA for attaching one 0.10 mm-thick film. From this observation, the dielectric-film-based waveguide possibly improves the field confinement at the interface between MPAs and films. Compared with the single-mode at 0.35 THz, the high-order mode at 0.68 THz exhibits different field distributions (see Fig. 7 (b)). The strongest field is located at the position of the middle (0.5 mm) when MPA integrated with a thick dielectric film (t=0.10 mm). Compared with that of the blank MPA, the integrated dielectric film raises lateral confinement to involve more fraction of field inside the MPA.

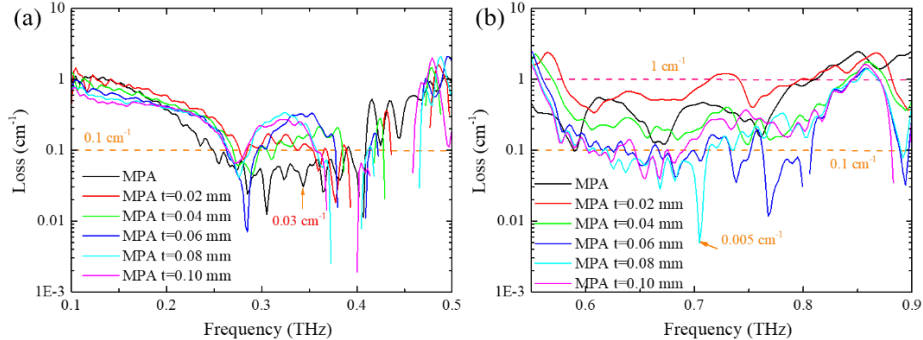

Fig. 8 The calculated propagation loss between 30- and 50- layered for blank MPA and various hybrid MPA with different dielectric film thickness.

Hybrid waveguides possess the ability to improves the trade-off between mode confinement and propagation loss. Therefore, we investigate the propagation loss of MPA-based hybrid waveguides. We calculate the loss between 30- and 50- layered of blank MPA and various hybrid MPA waveguides. As shown in Fig. 8 (a), for the low-frequency band of the hybrid waveguide, the dielectric film does not reduce the propagation loss of 0.35 THz. For example, compared with the blank MPA, the lowest loss of 0.35 THz is higher than 0.1 cm$^{-1}$ when the dielectric film thickness is 0.02 mm. For the high-frequency band above the Bragg bandgap, the propagation loss of 0.68 THz for 0.02 mm-thick film is larger than that of blank MPA. The 0.02 mm-thick film breaks the phase matching conditions and thus results in a weak confinement mode. With the increasing film thickness, the scattering loss becomes smaller. Thus, the loss of this hybrid waveguide can be optimized by changing the film thickness. The lowest propagation loss occurs at 0.70 THz for the hybrid waveguide of MPA integrated with a 0.08 mm-thick film, which is lower than 0.01 cm$^{-1}$. It notes that 0.08 mm-thick film prevents the field leaked out of MPA and thus raises the lateral confinement due to the primary mode field is trapped inside the MPA (see Fig. 6 (d)). From this observation, the dielectric-film-based ribbon waveguide possibly extends the guiding ability of MPA. The dielectric film-MPA hybrid waveguide modes perform stronger confinement and lower waveguide loss, which are beyond those properties in the pure MPA waveguide

### *3.2.2 Propagation loss dependent on the refractive index of the dielectric film*

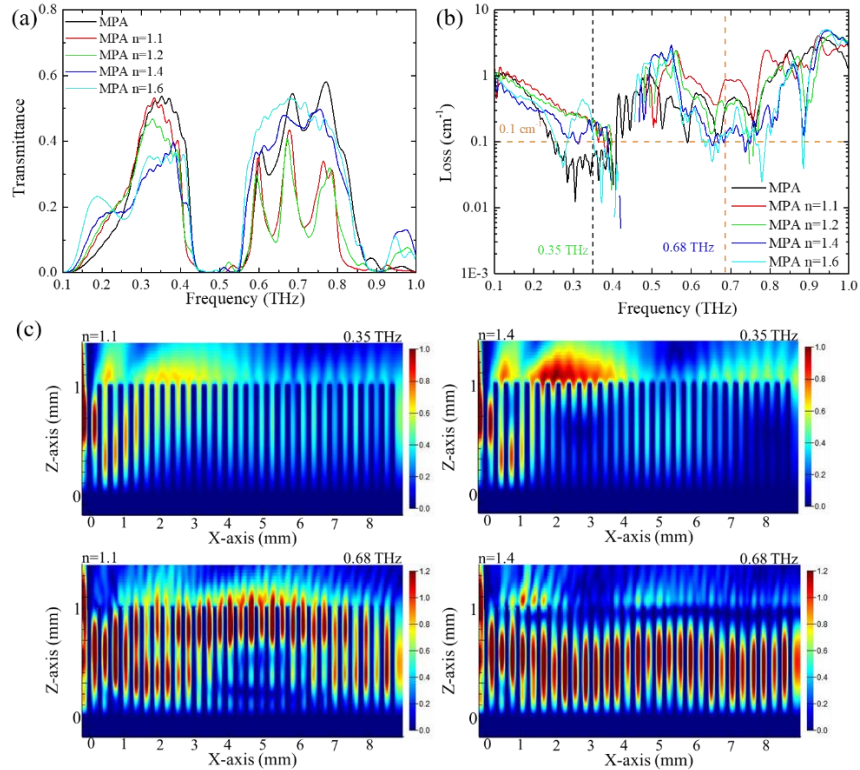

Fig. 9 (a) The calculated transmittance spectra of MPAs integrated with different thickness dielectric films. (b) The propagation loss of MPAs with different dielectric films. (c) electric field distribution at 0.35 THz and 0.68 THz for different film refractive indices, such as 1.1 and 1.4.

The refractive index of dielectric waveguide is also a giant factor for the guiding performance of hybrid waveguides. Therefore, in this section, we vary the refractive index of the dielectric film to investigate the loss and electromagnetic field distribution of MPA-based hybrid waveguide, where the film thickness is fixed as 0.10 mm. The loss of 0.10-thick film in FDTD simulation is 0.01 cm$^{-1}$. For the same waveguide scheme, the THz transmittance spectra for various refractive indices of a 0.10 mm-thick film on the MPA is investigated. Figure 9 (a) shows the dependence of the transmission spectra of MPA on film refractive indices. The transmittance spectra have been tailored by the dielectric film with various refractive indices. Compared with the blank MPA, the low refractive index film (n=1.1) on the MPA reduces the transmittance of the high-frequency band. With the increases film refractive index, the transmittance increases due to the improved confinement. We calculated the propagation loss of MPA with various dielectric film and depicted in Fig. 9 (b). The blank MPA achieves a loss at 0.35 THz lower than 0.1 cm$^{-1}$. By integrating a dielectric film on the MPA, the loss at 0.35 THz increased. For example, the loss is higher than 0.15 cm$^{-1}$ when the refractive index of the film is 1.1. Note that the film on the MPA results in a loosely confined mode. For the high-frequency band, the low refractive index film (n=1.1) on MPA shows a giant propagation loss at 0.68 THz, distinctly higher than that of blank MPAs. Increasing the refractive index of the dielectric film, the related loss can be reduced. Electric field distributions of 0.35 THz and 0.68 THz for MPAs with different films are given in Fig. 9 (c). At 0.35 THz, the dielectric film on MPAs guides more THz fields on the film surface and then increases the propagation loss. However, at 0.68 THz, the larger refractive index film improves the mode confinement. As shown in Fig. 9 (c), the most THz fields are confined in the MPA-based hybrid waveguides.

4. Conclusion

In summary, we have developed one hybrid waveguide consisting of photonic crystals based on metal pillar arrays (MPAs) and dielectric waveguides in 0.1-1 THz, which shows one apparent bandgap in 0.45–0.55 THz. Two resonance modes of MPAs can be termed as fundamental and high-order transverse magnetic (TM) modes. Both of them are sensitive to the changes in the surrounding environment. Results demonstrated that longer pillar is beneficial to reduce the propagation loss. For the fundamental mode in the low-frequency band, the lowest loss for 1.0 mm-MPAs is less than 0.1 cm$^{-1}$ at 0.35 THz. However, the propagation loss of the high-order mode at 0.68 THz is 0.2 cm$^{-1}$. To further reduce the propagation loss in MPAs, a dielectric film with low loss is integrated on the MPAs. By changing the dielectric film's thickness and refractive index, not only the propagation loss but also the field confinement can be optimized. The investigation shows the lowest loss is less than 0.1 cm$^{-1}$ at 0.68 THz because the high-order TM-mode THz waves are strictly confined inside the hybrid waveguide. The proposed MPA-based hybrid waveguides are possible to develop as terahertz integrated components such as filters, sensors, and splitters.

**Funding**

This study received funding from the Science and Technology Commission of Shanghai Municipality (No. 19590746000, No. 18590780100 and No. 17142200100), Innovation Program of Shanghai Municipal Education Commission (No. 2019-01-07-00-02-E00032).

**Authors contribution**

D. Liu organized and wrote the paper. All authors contributed to the paper.

**Disclosures**

The authors declare no conflicts of interest.